\documentclass[conference]{IEEEtran}
\IEEEoverridecommandlockouts
\usepackage{amsmath,graphicx}

\usepackage[utf8]{inputenc}
\usepackage[T1]{fontenc}
\usepackage{amsfonts,amssymb,mathtools}
\usepackage{bm}
\usepackage{url}
\usepackage{cite}
\usepackage{balance}

\usepackage[bb=pazo]{mathalpha}

\usepackage{graphicx}
\graphicspath{{./figs/}}
\usepackage{subfig}

\usepackage{comment}
\usepackage{algorithm}
\usepackage{algpseudocode}
\usepackage{xcolor}
\usepackage{soul}

\DeclareMathOperator*{\argmin}{arg\,min}

\title{Path Planning for 
Sound Speed Profile Estimation 
}

\author{Ludvig Lindström$^*$\thanks{This work was supported by the  Wallenberg AI, Autonomous Systems and Software Program (WASP) funded by the Knut and Alice Wallenberg Foundation. The first and second authors have contributed equally to the work.}, Tadas Paskevicius$^\dagger$, Andreas Jakobsson$^\dagger$, Isaac Skog$^{*,\ddag}$ \\ \\
$^*$Division of Communication Systems, Royal Institute of Technology, Sweden \\
$^\dagger$Centre for Mathematical Sciences, Mathematical Statistics, 
Lund University, Sweden\\
$^\ddag$FOI Swedish Defence Research Agency, Stockholm, Sweden 
}

\begin{document}

\maketitle

\begin{abstract}
Accurate estimation of the sound speed profile (SSP) is essential for underwater acoustic communication, sonar performance, and navigation, as the acoustic wave propagation depends strongly on the SSP. This work considers SSP estimation in a region of interest using an autonomous underwater vehicle (AUV) equipped with a conductivity-temperature-depth (CTD) sensor and an acoustic receiver measuring transmission loss (TL) from a sonar transmitter. The SSP is modeled using a linear basis-function expansion and is sequentially estimated with an unscented Kalman filter that fuses local CTD measurements with TL measurements. A receding-horizon path planning scheme is also employed to select future AUV positions by minimizing the predicted total sound speed variance. Simulations using the Bellhop acoustic wave propagation solver show that CTD measurements provide accurate local SSP estimates, whereas TL measurements are seen to capture the global characteristics of the SSP, with their joint use improving the reconstruction of both local variations and large-scale SSP behavior. The results also indicate that the proposed path planning strategy reduces the estimation uncertainty compared to constant-velocity motion, thereby enabling improved environmental characterization for underwater acoustic systems.

\end{abstract}

\section{Introduction}

Underwater acoustic systems play a critical role in a wide range of maritime applications, including underwater communication~\cite{stojanovic2009underwater,chitre2008recent}, surveillance~\cite{eleftherakis2020sensors}, navigation~\cite{kinsey2006survey}, and environmental monitoring~\cite{cauchy2023gliders}. To effectively design and operate these systems, it is essential to have good understanding of underwater acoustic propagation, which is generally characterized by its complex multipath nature~\cite{larsen2018acoustic,stojanovic2009underwater, chitre2008recent,YangLSMJ24_73}. The propagation depends on both the domain geometry and the sound speed profile (SSP). The SSP is often time-varying and spatially heterogeneous, especially in shallow and coastal waters, where it can exhibit significant variability due to surface heating and cooling cycles, wind-driven mixing, tidal currents, freshwater influx from river outlets, and bathymetric features~\cite{mousavi2025ocean, piao2023time, li2022novel, abraham2019underwater}. Since this variability can lead to significant degradation in system performance if not properly accounted for~\cite{piao2023time}, accurate estimation of the SSP are crucial for most maritime applications.
%
\begin{figure}[t!]
    \centering
    \includegraphics[width=1\linewidth]{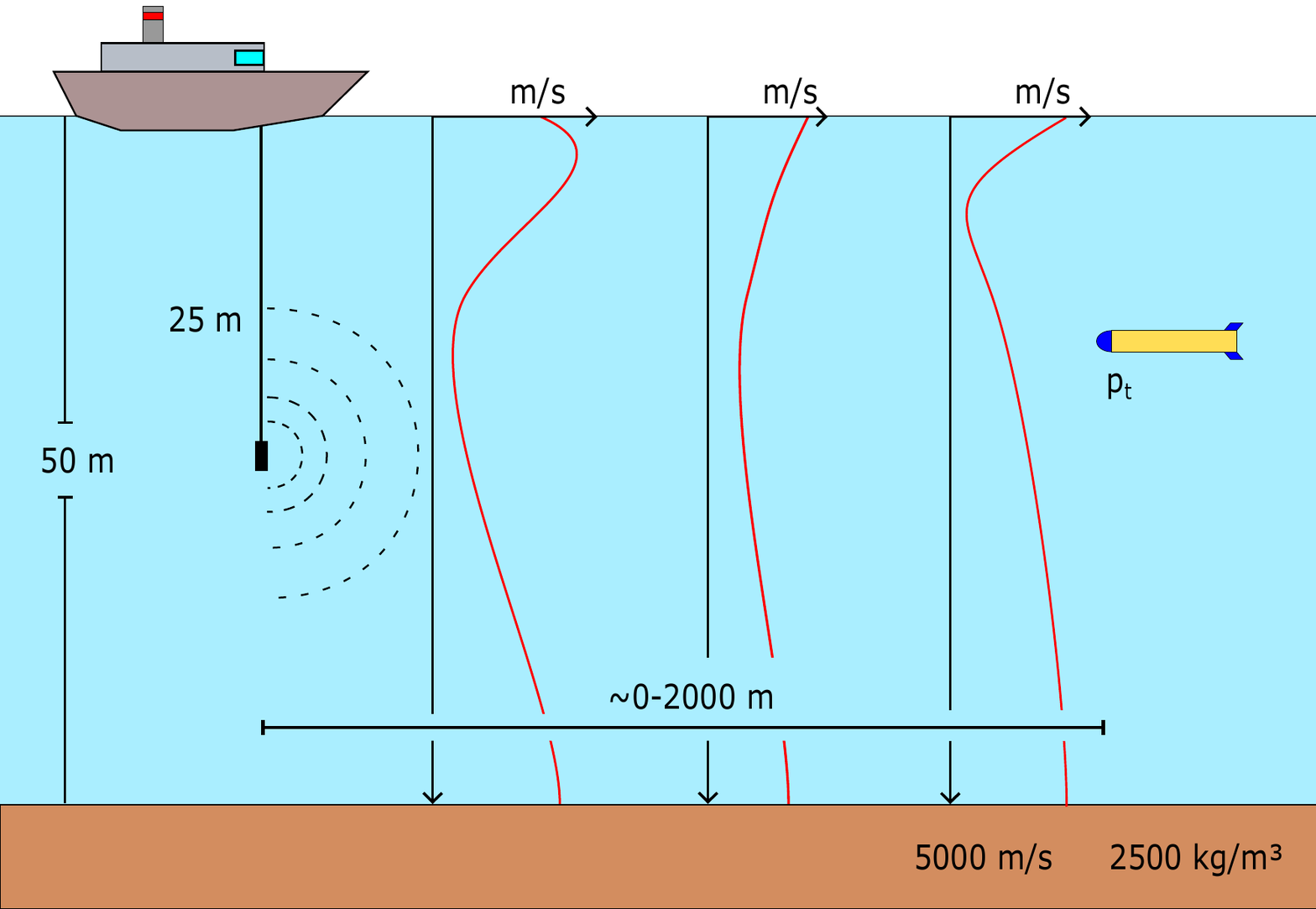}
    \caption{The studied system with an underwater transmitter and receiver at the boat and AUV, respectively. The SSP within the water volume is unknown and spatially varying (red lines).
    }
    \label{fig:systemFig}
\end{figure}
%
Currently, many SSP estimates are obtained via 
in-situ sound speed measurements, using e.g., conductivity-temperature-depth (CTD)  sensors, as well as large-scale estimates of the SSP derived from time-of-flight (ToF) measurements between coordinated transceiver pairs~\cite{mousavi2025ocean,radhakrishnan2024inversion}.
%
%
However, these approaches have limitations in spatial coverage, operational complexity, and resource requirements, thereby making large-scale SSP estimation challenging. 

This study proposes a novel method to estimate the SSP, utilizing sonar transmission loss (TL) measurements and in-situ CTD measurements collected by an autonomous underwater vehicle (AUV). Compared to conventional ToF-based SSP estimation methods, TL-based approaches generally require less coordination between transmitter and receiver pairs and, unlike CTD measurements, are affected by the global characteristics of the propagation channel. In this work, we formulate an unscented Kalman filter (UKF) that fuses, at a global scale, both TL and in-situ CTD measurements, demonstrating that this significantly improves the efficacy of the SSP estimate. Exploiting the resulting estimation procedure, we formulate a path planning scheme for the AUV
such that future locations are selected through an optimal control scheme attempting to minimize the global root-mean-square error (RMSE) of the SSP estimate.  
%
In summary, our contributions are:
\begin{itemize}
    \item Presenting a method for estimating the SSP using TL and in-situ CTD measurements.
    \item Showing that there is a significant gain using both TL and in-situ CTD measurements when estimating the SSP compared to only in-situ CTD measurements.
    \item Proposing a path planning scheme to minimize the estimation error in the resulting SSP estimate. 
\end{itemize}
In the interest of reproducibility, both the code and the datasets used in this paper are publicly available\footnote{{\em https://github.com/Tadaspas/Path-Planning-for-Sound-Speed-Profile-Estimation}}.


\section{Signal Model}\label{sec:Problem Formulation}

This work concerns the estimation of the SSP in a region of interest. To do so, an AUV is maneuvered in the vicinity of an active sonar, as illustrated in Fig.~\ref{fig:systemFig}. The AUV is assumed to be able to measure the sound speed at its current location and the TL from the sonar transmitter to the AUV. The latter also carries information about the SSP~\cite{qiao2019distributed}.

The in-situ CTD measurements collected by the AUV can, to the first-order, be modeled as 
\begin{equation} \label{eq:ctd_mesurment}
    y_t^{\text{\tiny L}} = c(\mathbf{p}_t;\boldsymbol{\theta}) + e^{\text{\tiny L}}_t, 
\end{equation}
where $c(\mathbf{p}_t;\boldsymbol{\theta})$ denotes a SSP model parameterized by the parameter $\boldsymbol{\theta}$, $\mathbf{p}_t$ the position of the AUV, and $e^{\text{\tiny L}}_t$ an additive measurement noise, here 
assumed to be well modeled as a Gaussian white noise with variance $\sigma_{\text{\tiny L}}^2$.
%
%
Furthermore, the TL measurements are assumed to be well described using an acoustic wave propagation, such that
%
%
\begin{equation} \label{eq:tl_mesurment}
    y_t^{\text{\tiny TL}} = f\Bigl({c}(\cdot; \boldsymbol{\theta}), {\mathbf p}_t\Bigr) +  e^{\text{\tiny TL}}_t,
\end{equation}
where $f(\cdot)$ denotes an acoustic wave propagation model   
and $e_t^{\text{\tiny TL}}$ is the measurement noise, here 
assumed to be well modeled as a Gaussian white noise with variance $\sigma_{\text{\tiny TL}}^2$. 

There are several models for describing the acoustic wave propagation, $f(\cdot)$ in \eqref{eq:tl_mesurment}, including Bellhop~\cite{porter2011bellhop} and KRAKEN~\cite{porter1992kraken}. Depending on the frequency band, these solve the wave equations in different ways, but all require a description of the environmental parameters, such as bottom properties, bottom topography, and the SSP. Here, to stress this, the dependency on the SSP, $c(\cdot;\boldsymbol{\theta})$, has been explicitly emphasized in \eqref{eq:tl_mesurment}; all other environmental parameters are assumed to be known.
%
%
%
%
%
%
There are several approaches to modeling the SSP~\cite{mousavi2025ocean,radhakrishnan2024inversion,zeng2025passive}. Here, 
we employ a linear basis-function expansion, such that
%
\begin{subequations}
\begin{equation} \label{eq:ssp_gp}
     c({\mathbf p}) = \boldsymbol{\phi}^\top({\mathbf p}) \boldsymbol{\theta},
\end{equation} 
where $(\cdot)^\top$ denoting the transpose, and
\begin{align}
    \boldsymbol{\theta} &= \left[ \begin{array}{cccc} \theta_0 & \theta_1 & \ldots & \theta_K \end{array}\right]^\top \\
     \boldsymbol{\phi}(\mathbf{p}) &= \left[ \begin{array}{cccc} 1 & \varphi_1(\mathbf{p}) & \ldots & \varphi_K(\mathbf{p}) \end{array} \right]^\top.
\end{align}
In this work, the basis functions are defined as
\begin{equation}
     \varphi_k({\mathbf p}) = \exp \Bigl( - \|{\mathbf p} - {\mathbf p}'_k\|^2_{\boldsymbol{\Lambda}_l}
     \Bigr),    
\end{equation}
\end{subequations}
with $\| {\mathbf x}\|^2_{\bf A} = {\mathbf x}^\top{\bf A} {\mathbf a}$, and where ${\mathbf p}'_k$ denotes the center point of the $k$:th basis function, and
$\boldsymbol{\Lambda}_l^{-1}$  the spread (length-scale) of the basis functions.



Assuming the SSP varies slowly over time, the models in \eqref{eq:ctd_mesurment} and \eqref{eq:tl_mesurment} may be used to define a state-space model for the complete system dynamics and measurements. To that end, let $\mathbf{y}_t=\begin{bmatrix}
 y_t^{\text{\tiny L}} & y_t^{\text{\tiny TL}}
\end{bmatrix}^\top$ be the measurement vector. Assuming the model parameters to vary according to a random walk,  the state-space model can be expressed as 
\begin{subequations}\label{eq:state_space}
\begin{align}
    \boldsymbol{\theta}_{t+1} &= \boldsymbol{\theta}_t + \mathbf{w}_t \\
    \mathbf{y}_t &= h(\boldsymbol{\theta}_t,\mathbf{p}_t) + {\bf e}_t
\end{align}
where
\begin{equation}
     h(\boldsymbol{\theta},\mathbf{p})=\begin{bmatrix} \boldsymbol{\phi}^\top({{\mathbf p}}) \boldsymbol{\theta} \\ f(c(\cdot\,; \boldsymbol{\theta}), {\mathbf p}) \end{bmatrix}\quad\text{and}\quad \mathbf{e}_t=\begin{bmatrix}
         e_t^{\text{\tiny L}} \\ e_t^{\text{\tiny TL}}
     \end{bmatrix}.
\end{equation}
Here, $\mathbf{w}_t$ denotes the process noise accounting for slow temporal variations in the SSP, which is assumed to be a white process with covariance $\bf{Q}$. Moreover, the measurement noise ${\bf e}_t$ has the covariance $\mathbf{R}=\text{diag}(\sigma^2_{\text{\tiny{L}}},\sigma^2_{\text{\tiny{TL}}})$.
\end{subequations}

\section{Sound Speed Profile Estimation}\label{sec:Proposed Solution}

Multiple nonlinear filtering methods exist for sequential estimation of the SSP parameters $\boldsymbol{\theta}_t$ in~\eqref{eq:state_space} from the measurement sequence $\mathbf{y}_{1:t}$ collected as the AUV traverses the underwater environment along the trajectory $\mathbf{p}_{1:t}$. Due to the highly nonlinear relationship between the SSP parameters $\boldsymbol{\theta}$ and the TL measurements $y_t^{\text{\tiny TL}}$, as described by the forward model $f(\cdot)$ in~\eqref{eq:tl_mesurment}, an UKF is employed in this work \cite{correnty2025sequential}.
Let $\hat{\boldsymbol{\theta}}_{t|t'}$ and $\boldsymbol{\Sigma}_{t|t'}$ denote the state estimate and the corresponding error covariance at time $t$, conditioned on measurements up to time $t'$. One iteration of the UKF for the state-space model in~\eqref{eq:state_space} is summarized in Alg.~\ref{alg:ukf}. Here, $\mathbb{S}(\cdot)$ denotes the sigma-point generation and $\mathbb{K}(\cdot)$ the Kalman measurement update; we refer the reader to~\cite{wan2000unscented} for details.

It is worth noting that the accuracy of the estimate $\hat{\boldsymbol{\theta}}_{t|t'}$ depends \emph{not only} on the measurement noise level but also on the {\em AUV positions} $\mathbf{p}_{1:t}$ at which the measurements are collected. This also implies that there are trajectories that offer better potential for improving the SSP estimate than others. To exploit this, we proceed to present a method for planning the AUV trajectory to reduce the uncertainty of the estimated SSP.

\begin{algorithm}[t]
\caption{Unscented Kalman filter}
\label{alg:ukf}
\begin{algorithmic}[1]
\State \textbf{Input:} $\mathbf{p}_t$, $\mathbf{y}_t$, $\hat{\boldsymbol{\theta}}_{t-1|t-1}$, $\boldsymbol{\Sigma}_{t-1|t-1}$, $\mathbf{Q}$, $\mathbf{R}$

\Statex State prediction

\State $\hat{\boldsymbol{\theta}}_{t|t-1} \gets \hat{\boldsymbol{\theta}}_{t-1|t-1}$ 

\Statex Covariance prediction
\State $\boldsymbol{\Sigma}_{t|t-1} \gets \boldsymbol{\Sigma}_{t-1|t-1} + \mathbf{Q}$ 

\Statex Generate sigma points
\State $\{\mathcal{X}^{(i)}_{t|t-1}\}_{i=0}^{2n} 
\gets \mathbb{S}\!\left(\hat{\boldsymbol{\theta}}_{t|t-1}, 
\boldsymbol{\Sigma}_{t|t-1}\right)$ 

\For{$i = 0, \ldots, 2n$}
\Statex Propagate sigma points through measurement model
    \State $\mathcal{Y}^{(i)} 
    \gets h\!\left(\mathcal{X}^{(i)}_{t|t-1}, \mathbf{p}_t\right)$ 
\EndFor

\Statex Measurement update
\State $\hat{\boldsymbol{\theta}}_{t|t}, 
\boldsymbol{\Sigma}_{t|t}
\gets \mathbb{K}\!\left(
\{\mathcal{X}^{(i)}_{t|t-1}\},
\{\mathcal{Y}^{(i)}\},
\mathbf{y}_t,
\mathbf{R}
\right)$ 

\State \textbf{return} $\hat{\boldsymbol{\theta}}_{t|t}$, $\boldsymbol{\Sigma}_{t|t}$
\end{algorithmic}
\end{algorithm}

\section{Path Planning}\label{sec:Path Planning}

In this section, we examine how the SSP estimate can be improved by controlling the AUV's motion using a receding-horizon path-planning strategy. This means that, at each time step, the AUV evaluates all feasible trajectories over a horizon of $T$ samples and selects the trajectory that minimizes a measure of the predicted uncertainty of the SSP. The AUV then moves one step along said path. After the next measurement, the state estimate is updated, and all the realizable paths are again considered using the updated estimate. The optimization problem to be solved in each step may thus be expressed as
\begin{algorithm}[t]
\caption{Receding-Horizon Path Planning Step}
\label{alg:pathPlan}
\begin{algorithmic}[1]
\State \textbf{Input:} $\mathbf{p}_t$, $\hat{\boldsymbol{\theta}}_{t|t}$, $\boldsymbol{\Sigma}_{t|t}$, $T$, $\mathbf{R}$, $\mathbf{Q}$

\Statex Compute optimal trajectory as using \eqref{eq:minL}
\State $\mathbf{p}_{t+1:t+T}^\ast\gets\argmin\mathcal{L}(\mathbf{p}_{t+1:t+T})$

\Statex Propagate AUV position
\State $\mathbf{p}_{t+1} \gets \mathbf{p}_{t+1}^{\ast}$ 

\Statex Collect measurements
\State $\mathbf{y}_{t+1}\gets \text{Measure}(\mathbf{p}_{t+1})$ 

\Statex Update estimate using using Alg.~\ref{alg:ukf}
\State $\hat{\boldsymbol{\theta}}_{t+1|t+1},\boldsymbol{\Sigma}_{t+1|t+1}\gets \text{UKF}(\mathbf{p}_{t+1},\mathbf{y}_{t+1},\hat{\boldsymbol{\theta}}_{t|t},\boldsymbol{\Sigma}_{t|t},\mathbf{R},\mathbf{Q})$ 

\State \textbf{return} $\mathbf{p}_{t+1}$, $\hat{\boldsymbol{\theta}}_{t+1|t+1}$, $\boldsymbol{\Sigma}_{t+1|t+1}$
\end{algorithmic}
\end{algorithm}
\begin{equation}\label{eq:minL}
    \mathbf{p}_{t+1:t+T}^\ast=\argmin_{\mathbf{p}_{t+1:t+T}\in\Omega_T}\mathcal{L}(\mathbf{p}_{t+1:t+T}),
\end{equation} 
where $\Omega_T$ is the set of all realizable paths over the time-horizon $T$. The objective function in \eqref{eq:minL} is here defined as 
\begin{equation}
    \mathcal{L}(\mathbf{p}_{t+1:t+T})=\sum_{i=1}^{T} \lambda^{i} \sigma^2_{\text{tot},i}(\mathbf{p}_{t+1:t+i}),
\end{equation}
where $\sigma^2_{\text{tot}, i}(\mathbf{p}_{t+1:t+i})$ denotes the predicted total sound speed variance after $i$ future measurements, with $0<\lambda\leq 1$ denoting a discount factor that places greater emphasis on earlier measurements. This reflects a setup in which the SSP estimate is updated after each measurement, making long-term predictions unreliable, implying that the planner should prioritize early decreases in the total variance rather than the minimum over the whole planning horizon. The predicted total variance is~\cite{kay1993statistical} 
\begin{equation}\label{eq:xi}
\begin{split}
    \sigma^2_{\text{tot},i} =&\int_{V'} \boldsymbol{\phi}^\top(\mathbf{p}) \boldsymbol{\Sigma}_{t+i|t} \boldsymbol{\phi}(\mathbf{p}) d\mathbf{p},
\end{split}
\end{equation}
where $V'$ is the region of interest and $\boldsymbol{\Sigma}_{t+i|t}$ is the predicted covariance of the SSP at time step $t+i$. In the interest of notational brevity, the explicit dependence on the sequence of positions $\mathbf{p}_{t+1:t+i}$ is dropped from the total variance $\sigma^2_{\text{tot},i}$, and the predicted covariance $\boldsymbol{\Sigma}_{t+i|t}$. The numerical minimization of~\eqref{eq:minL} is here implemented using differential evolution, as detailed in~\cite{qin2008differential}, with a population size of 20, a generation count of 50, and mutation and crossover factors of 0.7 and 0.9, respectively.

\subsubsection{Efficient Prediction of State Covariance}


The total variance for the potential paths the AUV can take is given by the $i$-step prediction of the state covariance matrix, $\boldsymbol{\Sigma}_{t+i|t}$.
%
%
The covariance can be computed, for example, by sequentially updating the state covariance matrix estimate using the UKF at each position along the path. 
As this requires multiple UKF predictions along each potential path, such an approach quickly becomes computationally prohibitive, especially for larger $T$. As an alternative, we here use the summed Fisher information to calculate $\boldsymbol{\Sigma}_{t+i|t}$, i.e., expressing $\boldsymbol{\Sigma}^{-1}_{t+i|t}$ as~\cite{hu2025target}
\begin{subequations}
\begin{equation}
    \boldsymbol{\Sigma}^{-1}_{t+i|t}= \boldsymbol{\Sigma}_{t|t}^{-1}+\sum_{j=1}^{i}{\mathbf{H}}_t({\mathbf p}_{t+j}) {\mathbf R}^{-1} {\mathbf{H}}_t^\top({\mathbf p}_{t+j}),
\end{equation}
where
\begin{align}
{\bf H}_t({\mathbf p})=&\left.\frac{\partial h(\boldsymbol{\theta},{\mathbf p})}{\partial \boldsymbol{\theta}}\right|_{\boldsymbol{\theta}=\hat{\boldsymbol{\theta}}_{t|t}}
\end{align}
\end{subequations}
denotes the gradient of the measurement function with respect to $\boldsymbol{\theta}_{t|t}$.
which may be computed as a numerical derivative using the Bellhop model.

Assuming that the SSP is time-invariant in the planning horizon, ${\mathbf{H}}_t({\mathbf p})$
can be computed using 
bilinear interpolation on a grid spanning the planning horizon box. This is computationally preferable, as adding measurement points is less computationally demanding than executing a Bellhop simulation at each spatial position.  

\begin{figure}[t]
\centering
\includegraphics[width=0.9\linewidth]{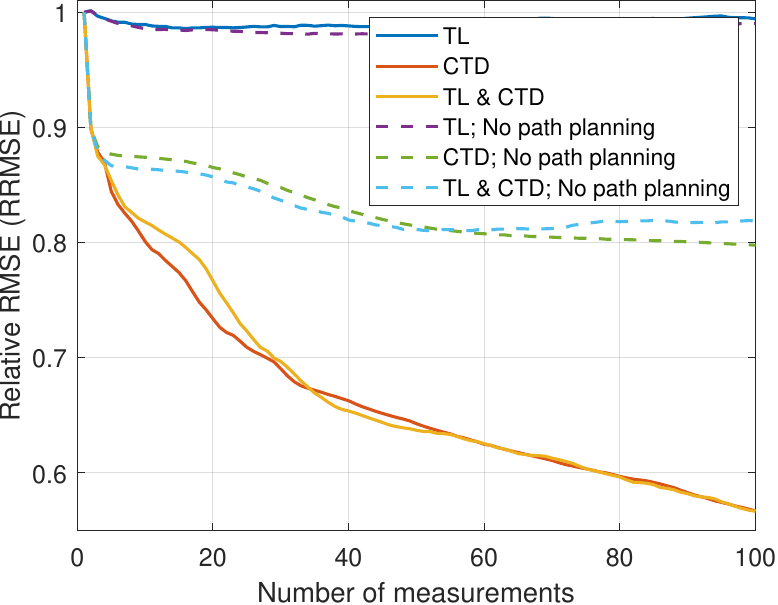}
\caption{The RRMSE versus the number of measurements.
}
\label{fig:multi_RMSE_vs_N_big}
\end{figure}

\begin{table}[t]
\caption{Simulation parameters.}
    \centering
    \begin{tabular}{l|c}
      \hline
      \hline
        Water depth & $50$ m    \\ 
        \hline
        Transmission range & $2000$~m \\
        \hline
        Number of rows of basis function & $6$
        \\
        \hline
        Number of columns of basis function & $6$
        \\
        \hline
        Spread of the basis function in depth ($[\mathbf{\Lambda}^{-1}_{l}]_{1,1}$) &  $(50/6)^2$~m$^2$
        \\
        \hline
        Spread of the basis function in range ($[\mathbf{\Lambda}^{-1}_l]_{2,2}$) &  $(2000/6)^2$~m$^2$
        \\
        \hline
        Transmitter depth & $25$~m \\ 
        \hline
        Transmitted frequency & $5000$~Hz\\
         \hline
         Bottom sound speed & $5000$ m/s  \\
         \hline
         Bottom density &  $2500$ kg/m$^3$ \\
         \hline
         Sound speed measurement std ($\sigma_{\text{L}}$) & $10^{-2}$~m/s \\
         \hline
         Transmission loss std ($\sigma_{\text{TL}}$)  & $10^{-5}$~Pa 
         \\
         \hline
         Process noise variance ($\bf{Q}$) & $10^{-3}\mathbf{1}$~m/s 
         \\
         \hline
         Initial AUV position ($\mathbf{p}_0$)  & $2000,15$ m \\
         \hline
         Initial estimate ($\hat{\boldsymbol{\theta}}_{0|0}$) &      $\begin{bmatrix}
    1500&\mathbf{0}^\top    \end{bmatrix}^\top~\text{m/s}$ 
         \\
         \hline
         Initial estimate variance ($\boldsymbol{\Sigma}_{0|0}$) & $5^2\mathbf{I}$~m$^2$/s$^2$\\
         \hline
         \hline
    \end{tabular}
    
    \label{tab:parameter values}
\end{table}

\subsubsection{Path Constraints From AUV Motion Model}

 In this work, we assume that the AUV motion follows a three dimensional extension of the bicycle kinematic model~\cite{kong2015kinematic}, wherein the AUV state vector consists of the position ${\mathbf p}(t) \in \mathbb{R}^3$, yaw heading angle $\psi(t)$, pitch angle $\theta(t)$, and speed $v(t)$, i.e.,
\begin{equation}
\mathbf{x}_c^{auv}(t) = \begin{bmatrix} \mathbf{p}^\top(t) & v(t) & \psi(t) & \theta(t) \end{bmatrix}^\top,
\end{equation}
with continuous-time dynamics
\begin{subequations}\label{eq:cont_dyn}
\begin{align}
    \dot{\mathbf{p}}(t) &= v(t)\mathbf{M}(\psi(t),\theta(t))\hat{\mathbf{e}}_x\\
    \dot{\psi}(t) &= v(t)\tan(\delta_{\psi}(t))/L \\
    \dot{\theta}(t) &= v(t)\tan(\delta_{\theta}(t))/L \\
    \dot{v}(t) &= a(t).
\end{align}
\end{subequations}
Here, $\mathbf{M}(\psi(t),\theta(t))$ denotes a three dimensional rotation matrix, $\hat{\mathbf{e}}_x$ a unit vector, $L$ a length scale analogous to the wheelbase in the bicycle model, and the acceleration, yaw, and pitch steering make up the action vector $\mathbf{u}(t)=\begin{bmatrix}
    a(t) & \delta_{\psi}(t) &\delta_{\theta}(t)
\end{bmatrix}^\top$. The state and action vector are restricted such that
\begin{subequations}\label{eq:conditions}
\begin{align}
    &|\delta_{\psi}(t)| \leq \delta_{\max},\\
    &|\delta_{\theta}(t)| \leq \delta_{\max},\\
    &|a(t)| \leq a_{\max},\\
    &v_{\min} \leq v(t) \leq v_{\max}\\
    &\mathbf{p}(t)\in V',
\end{align}
\end{subequations}
 where the maximum yaw and pitch steering angles are $\delta_{\max}$, the maximum acceleration is $a_{\max}$, and the maximum and minimum speeds are $v_{\min}$ and $v_{\max}$, respectively.
%
%
Given a sampling period $\Delta t$, let 
$\mathbf{x}_k = \mathbf{x}_c^{auv}(k\Delta t)$ and $\mathbf{p}_k$ the
corresponding position component.
The set of dynamically realizable paths
$\mathbf{p}_{1:T}$ are then defined as
\begin{align}
\Omega_T \triangleq 
\Big\{ \mathbf{p}_{1:T} \;\Big|\; 
&\exists \, \mathbf{u}(\cdot)
\text{ s.t. eq. \eqref{eq:cont_dyn} and \eqref{eq:conditions} are satisfied,} \nonumber\\
& \mathbf{p}_k = \mathbf{p}(k\Delta t), \; k=1,\dots,T
\Big\}.
\end{align}
Furthermore, we let the action vector be parameterized using a quadratic Bezier curve, i.e.,
\begin{subequations}
\begin{align}
\delta_{\psi} &= (1-\tau)^2 b_{\psi 0} + 2(1-\tau)\tau b_{\psi 1} + \tau^2 b_{\psi 2} \\
\delta_{\theta} &= (1-\tau)^2 b_{\theta 0} + 2(1-\tau)\tau b_{\theta 1} + \tau^2 b_{\theta 2},
\end{align}
\end{subequations}
where $b_0$, $b_1$, and $b_2$ are the Bezier control points for $\delta_{\psi}$ and $\delta_{\theta}$, and $\tau = {t}/(T\Delta t) \in [0, 1]$ is the normalized time parameter. Imposing the maximum yaw and pitch on the steering angles, the parametrized values are then  restricted as
\begin{subequations}
\begin{align}
b_0 &= \delta_{\text{prev}} \\
|b_1|&\leq\delta_{\max} \\
|b_2| &\leq \delta_{\max}
\end{align}
\end{subequations}
where $\delta_{\text{prev}}$ is the previous executed steering angle.

\begin{table}[t]
\caption{Motion parameters.}
    \centering
    \begin{tabular}{l|c}
      \hline
      \hline
        Sampling period ($\Delta T$) & $2.5$~s    \\ 
        \hline
        Planning horizon ($T$) & 20 steps \\
        \hline
        Discount factor ($\lambda$) & $0.95$ \\ 
        \hline
        Bicycle length scale ($L$) & $25$~m\\
        \hline
        Maximum steering angle ($\delta_{\max}$) & $10^\circ$  \\
         \hline
         Maximum acceleration ($a_{\max}$) & 0~m/s$^2$\\
         \hline
          Starting speed & $2$~m/s\\
          \hline
         Starting depth & $15$~m\\
          \hline
         Starting range & $2000$~m\\
         \hline
         \hline
    \end{tabular}
    
    \label{tab:motion values}
\end{table}

\section{Numerical examples}


A Monte-Carlo simulation using the Bellhop wave propagation model~\cite{porter2011bellhop} was conducted to evaluate the performance of the proposed SSP estimation method and path-planning approach. A system setup similar to that shown in Fig.~\ref{fig:systemFig} was simulated. To reduce the computational complexity, the simulations were made only in two dimensions. Further, the system parameters in Tab.~\ref{tab:parameter values} were used, with the $6\times 6$ basis functions being evenly distributed over the $2000\times50$~m$^2$ area of interest.
Here, both the bottom density and sound propagation speed in the bottom are assumed to be known.
%
%
%
Finally, the AUV's motion parameters are given in Tab.~\ref{tab:motion values}.
%
%

A measure of the method's performance is the RMSE,
\begin{equation}
    \text{RMSE}^{[m]}_t=\sqrt{\int_{V'}{(c(\mathbf{p},\hat{\boldsymbol{\theta}}^{[m]}_t)-c(\mathbf{p},\boldsymbol{\theta}^{[m]})})^2d\mathbf{p}},
\end{equation}
where $m$ indicates the $m$:th Monte-Carlo realization. If the estimator $\hat{\theta}$ is unbiased, the total variance and the square of the RMSE equate \cite{kay1993statistical}. In this paper, the performance metric used is the relative RMSE (RRMSE)
\begin{equation}
    \text{RRMSE}_t^{[m]}=\frac{\text{RMSE}^{[m]}_t}{\text{RMSE}^{[m]}_0}.
\end{equation}
Two types of steering protocols were considered, one in which the AUV moves without path planning with constant velocity and constant depth towards the transmitter, and the other in which the motion of the AUV optimizes the cost function in~\eqref{eq:minL}. 
 For each steering protocol, three sensor configurations were evaluated. One, where the AUV only collected TL measurements, another where the AUV only collected CTD measurements, and finally one where the AUV collected both TL and CTD measurements jointly. 
\begin{figure*}
    \centering
    \subfloat[True SSP.]{\includegraphics[width=0.48\linewidth]{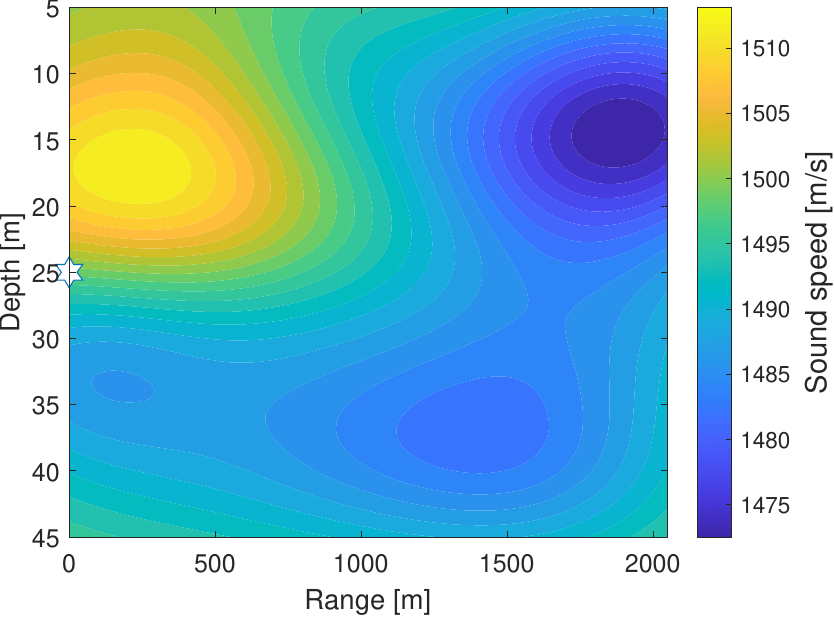}\label{fig:example_true_field}}
    \subfloat[SSP using TL measurements. The RRMSE\\ is here $0.9268$.]{    \includegraphics[width=0.48\linewidth]{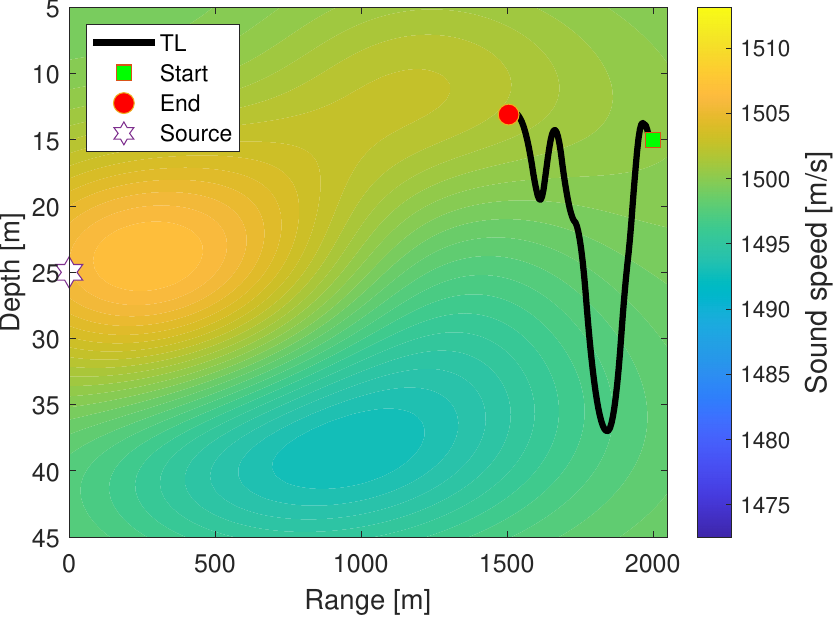}\label{fig:example_field_tl}}\\
    \subfloat[SSP using CTD measurements. The  RRMSE\\ is here $0.4067$.]{    \includegraphics[width=0.48\linewidth]{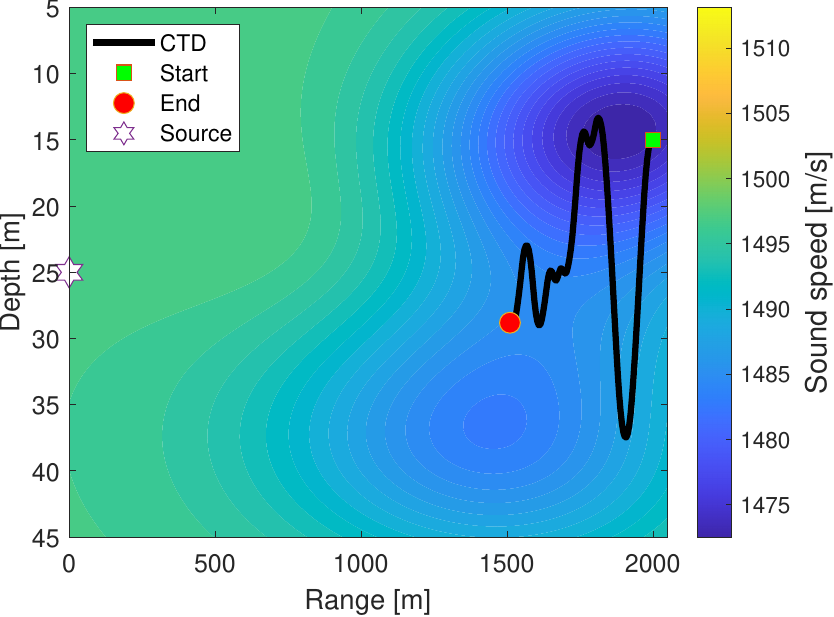}\label{fig:example_field_ctd}}
    \subfloat[SSP using TL and CTD measurements. The RRMSE\\ is here $0.4095$.]{    \includegraphics[width=0.48\linewidth]{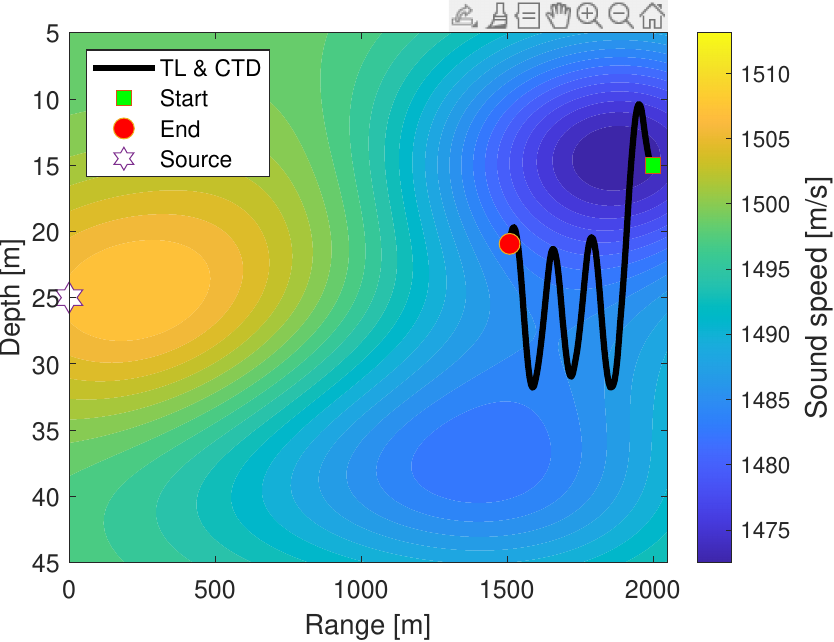}\label{fig:example_field_tl_ctd}}
    \caption{An example of the sound speed estimated after 100 measurements using TL, CTD, as well as with CTD and TL jointly,
    as well as each of the selected AUV paths using the proposed scheme.
    }
    \label{fig:ssp_fields}
\end{figure*}
To show the efficiency of the motion models and sensor configurations, 350 Monte-Carlo realizations were formed, where for each realization the true sound speed parameters $\boldsymbol{\theta}\sim \mathcal{N}( \hat{\boldsymbol{\theta}}_{0|0},\boldsymbol{\Sigma}_{0|0})$ were drawn anew.
Fig.~\ref{fig:multi_RMSE_vs_N_big} shows how the average RRMSE over the Monte-Carlo simulations varies as a function of the number of measurements. As seen in the figure, the introduction of path planning notably improves the SSP estimates when using CTD measurements. However, comparing the configurations using only CTD measurements and the configurations using both TL and CTD measurements, no notable improvement to the RRMSE is seen.  

\begin{figure}[t]
\centering
\includegraphics[width=0.9\linewidth]{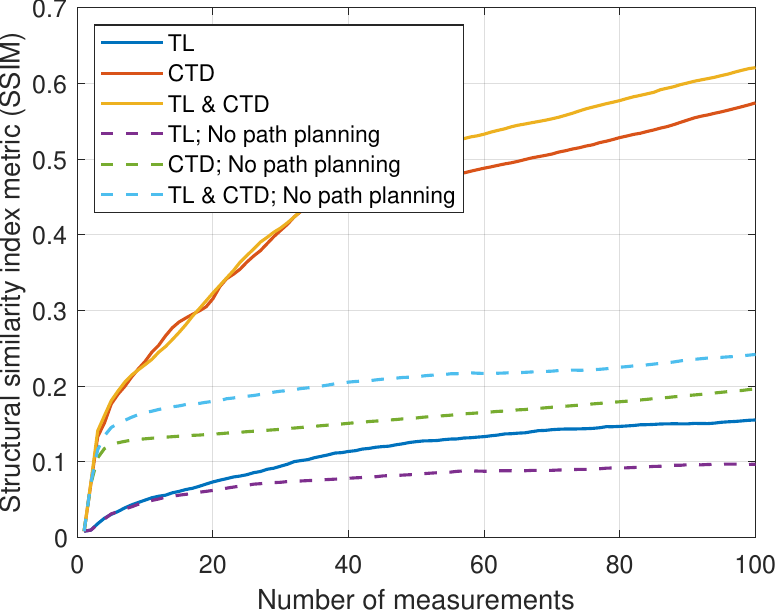}
\caption{The average structural similarity index metric versus the number of measurements.
}
\label{fig:final_ssim}
\end{figure}

To illustrate the efficacy of the TL methods, consider the examples illustrated in Fig.~\ref{fig:ssp_fields}. Here, Fig.~\ref{fig:example_true_field} shows the true SSP, 
whereas the estimated SSP for each measurement configuration, as well as the path taken by the AUV, are presented in Figs.~\ref{fig:example_field_tl}-\ref{fig:example_field_tl_ctd}.  Studying the three methods, it is clear that when using only TL measurements, as shown in Fig~\ref{fig:example_field_tl}, some of the SSP features are identified, such as the increase in sound speed near the transmitter. Still, it struggles to capture the position and amplitude of these features. Meanwhile, using the CTD measurements, shown in Fig~\ref{fig:example_field_ctd}, the SSP is well estimated locally but not at a distance from the AUV. When using the TL and CTD measurements jointly, as shown in Fig~\ref{fig:example_field_tl_ctd}, 
one can accurately determine the SSP for the entire region of interest, although it is worth noting that the precise shape and position of the high-velocity region in the upper left corner is not entirely accurately positioned. Notably, when examining the corresponding  RRMSE for the case using only TL measurements, the scores are only marginally better than not doing an estimation, whereas 
the estimate using the joint measurements scores worse than when only using CTD measurements, clearly illustrating the shortcomings of using the RRMSE as a metric for this problem.  As an alternative, we instead consider the structural similarity index metric (SSIM), as discussed in~\cite{nilsson2020understanding}, which compares the intensity, contrast, and structure of the estimated and true SSP. 
In contrast to the RRMSE results, Fig.~\ref{fig:final_ssim} illustrates that using joint TL and CTD measurements captures the behavior of the SSP better than using only CTD measurements. It also shows the importance of path planning, even when using only TL measurements, indicating that this is a more suitable metric than RRMSE for the problem at hand.


\section{Conclusions}

The numerical results show that TL and CTD measurements provide complementary information for SSP estimation. CTD measurements yield accurate local estimates of the sound speed, whereas TL measurements capture global characteristics of the acoustic propagation environment. Their joint use improves the reconstruction of both local SSP variations and large-scale structural behavior of the SSP compared to using either measurement type alone.
The receding-horizon path planning scheme reduces the estimation uncertainty by selecting future measurement locations that minimize the predicted SSP variance. This improves performance compared to a constant-velocity motion strategy, particularly when relying on in-situ CTD measurements. Furthermore, RRMSE alone does not fully reflect spatial similarities between the true and estimated SSP, which structural metrics such as SSIM better capture.
In a broader perspective, these findings indicate that combining acoustic propagation measurements with platform path-planning enables efficient acoustic propagation characterization over large areas. This has a direct implication for underwater communication, sonar performance prediction, and navigation, where accurate knowledge of the acoustic propagation environment is critical.

\bibliographystyle{IEEEtran}
\bibliography{refs} 




\end{document}